# The impact of moving expenses on social segregation: a simulation with RL and ABM

Xinyu Li
University of Washington

**Abstract**

Over the past decades, breakthroughs such as Reinforcement Learning (RL) and Agent-based modeling (ABM) have made simulations of economic models feasible. Recently, there has been increasing interest in applying ABM to study the impact of residential preferences on neighborhood segregation in the Schelling Segregation Model. In this paper, RL is combined with ABM to simulate a modified Schelling Segregation model, which incorporates moving expenses as an input parameter. In particular, deep Q network (DQN) is adopted as RL agents' learning algorithm to simulate the behaviors of households and their preferences. This paper studies the impact of moving expenses on the overall segregation pattern and its role in social integration. A more comprehensive simulation of the segregation model is built for policymakers to forecast the potential consequences of their policies.

## 1. Introduction

Simulations of economic models help policymakers to project the consequences of particular political decisions[1]. For example, policymakers can use a simulation of the Schelling model to investigate the impact of migration on the segregation behavior of the existing populations in events such as the influx of refugees into Europe in 2018 [2]. The Schelling model lays a theoretical foundation for viewing residential preferences as relevant to ethnic patterns [5]. While the mathematical basis for the classic Schelling model has been well-studied by prior research [3], advances in computing hardware and algorithms have made it feasible for researchers to simulate the segregative outcomes under specific residential preferences and further understand the underlying social dynamics [4,5]. For example, Fossett explores the role of social distance and ethnic preference in residential segregation by using the SimSeg computational model [6]; ASSURE, a methodology that can simulate urban growth and intra-urban social segregation, is proposed [7] to simulate urban growth of and social segregation in Kampala (Uganda). ABM is one of the tools widely used to simulate the behaviors of agents, but it alone is not sufficient for agents to learn the "optimal" behavior and the "balance" between exploration and exploitation. These two challenges can be overcome by recent advances of RL algorithms in finding optimal policies [8] and achieving a good balance between exploration and exploitation [9]. In this paper, RL is combined with ABM to simulate a modified Schelling model and study the impact of moving expenses on the residential outcome. The remainder of the paper is organized as follows: Section 2 provides useful background information and definitions of key algorithms, Section 3 presents the methodology and modeling process, Section 4 discusses experiment results, and Section 5 concludes the paper.

## 2. Background

### 2.1 Agent-Based Modeling

Traditional economics models are often based on mathematical equations that are derived from assumptions that may have been made for the convenience of analytical tractability [11]. One prominent feature that makes ABM different from traditional models is that it generalizes these assumptions [11] and thus provides insight into a broader behavior pattern. Additionally, ABM specified the properties, states, and behaviors of each agent, which is an autonomous element of a simulation [12]. ABM encodes [12] each agent by the rules of the environment and records its interactions with other agents. The dynamic behavior of heterogeneous agents is represented by both rule-based and analytical decision-making functions [13]. ABM has been widely applied to simulate many economic models, including the Schelling Segregation model.

### 2.2 The Schelling Segregation Model

The Schelling Segregation Model was first proposed by Thomas Schelling in 1971 to study the impact of residential preferences on neighborhood segregation [14]. It suggests that minor variance in neighbor preferences can cause aggregate results [15] of social segregation. In a bounded neighborhood model, every household (agent) has a preferred combination of its neighbors. Based on its tolerance level of neighbors from different ethnic groups, the agent can choose to stay in the area or move to other locations. Since its proposal, many studies have examined and extended the classic Schelling Segregation Model. Jensen etc. [16] explored the effect of altruistic agents in the model and found that even a minor proportion of altruists can have dramatic catalytic effects on the collective utility of the system. Jani [17] investigated the impact of resource scarcity on intolerance towards out-group members and redefined the model as a zero-sum game. He observed that a decline in environmental resources can lead to a high degree of intolerance towards out-group members, which can eventually lead to social disintegration. Silver etc. [18] extended the model by incorporating urban venues as a model input and used ABM to simulate the relationship between urban venues and social segregation.

### 2.3 Reinforcement Learning

Reinforcement Learning is a learning algorithm that trains agents to achieve maximized reward [19]. An RL agent learns to exploit actions that yielded high rewards in the past while exploring other possible actions to achieve higher rewards in the future. To do so, the agent learns to interact with the environment by finding actions that return the maximum future reward through trial-and-error and a delayed reward mechanism [19]. The future reward at time t is defined [20] as

$$R_t = \sum_{t'=t}^{T} \gamma^{(t'-t)} r(s_{t'}, a_{t'})$$

where $\gamma$ is a discounting factor, s is the state, and a is the action. The optimal action-value function [20] is the maximum expected reward $Q^*(s, a) = max_\pi E[R_t|s_t = s, a_t = a, \pi]$, where $\pi$ is a policy mapping sequences to actions. According to the Bellman equation, if the optimal value of $Q^*$ at the next state was known for all possible actions, then the optimal action-value function becomes $Q^*(s, a) = E_{s'\sim\varepsilon}[r + \gamma max_{a'} Q^*(s', a')|s, a]$. The RL algorithm used in this paper is Deep Q learning (DQN), which combines a convolutional neural network with Q-learning algorithm [20]. In recent years, DQN has been widely applied in robot motion planning tasks due to its superior performance [21] in perceiving environments with large state space in high dimensions.

## 3. Methods

### 3.1 States and Actions

There are $3^{n^2}$ available states in the environment and each agent has an observation window of size $n \times n$. Based on the number and type of agents in its observation window, the agent can decide which actions to take among the five available options: it can choose to move up, down, left, right, or stay still.

### 3.2 Age

All agents take one action per iteration at a random sequence. There are 80 iterations for each agent, which represent the resident's age. After an agent reaches the maximum iteration, it "dies" and is replaced by a new agent at a random location.

### 3.3 Neural Networks Environment

The simulation adopts the initial experiment setting in [10] where two types of agents (type A and type B) interact on a two-dimensional grid of size $50 \times 50$. Each location can have three possible values: it is labeled as 1 for agents of the same type, 0 for empty space, and -1 for agents of different types. In order to create a multi-agent RL environment, two independent neural networks are used for two types of agents respectively. Finally, each agent strives to maximize its reward function

$$Q^* = max_\Phi E[\sum_{t=0}^{80} \sum_{i=1}^{N} \gamma^t r_t^i | s_t^i ]$$

where $\phi$ denotes the Q-Network, N denotes the number of agents of the same type, and s denotes the state that contains both the spatial observation and age. The parameters used to initialize each network are listed in Table 1.

| Parameter | Value |
| --- | --- |
| Discount factor γ | 0.99 |
| Batch size | 256 |
| Learning rate | 0.001 |
| Number of iterations | 3000 |
| Number of training steps | 3000 |
| Optimizer | Adam [22] |
| Replay buffer size | $10^6$ |
| Initial exploration rate | 0.9 |
| Final exploration rate | 0 |
| Exploration decay | $10^5$ |

Table 1. Parameters used to initialize Deep Q-Networks

### 3.4 Rewards

- **Tolerance reward** Each agent is randomly assigned with a tolerance level $\alpha$: $\alpha = 0$ represents high tolerance towards neighbors from different ethnic groups, $\alpha = 0.5$ represents medium tolerance, and $\alpha = 1$ represents high intolerance. To better simulate residential preferences, an agent will receive a reward of $a - \alpha b$, where $a$ and $b$ denote agents of type A and type B respectively.

- **Interdependence reward** Agents are encouraged to interact with one another and will receive a penalty of -1 if they choose to stay still. Moreover, if an agent moves to a location that another agent of a different type occupies, it can "kill" that agent and occupy the place. In return, the "victim" agent will receive a penalty of -1.

- **Survival reward** The aim of a survival reward is for agents to learn that dying should be avoided. Therefore, they are awarded +0.1 for every iteration they survive and receive a penalty of -1 if they die.

### 3.5 Moving expenses

To better simulate the cost of moving in real life, a moving expense parameter is introduced. Everytime an agent takes an action, there is an associated moving expenses $c$ with three possible values: $0.3$ represents low moving expenses, $0.6$ represents medium moving expenses, and $0.9$ represents high moving expenses.

## 4. Results

The experiment is simulated with varying levels of tolerance α and moving expenses c. Figure 1 shows the residential outcomes over 1000 iterations. The scales are adjusted for better visualization of the segregative patterns. While high intolerance level encourages households to move away from neighbors of different races, there is an increasing trend in social integration as the moving costs increase. In particular, the bottom right plot demonstrates a pattern where households distribute evenly along the diagonal line under high moving expenses, despite the almost equivalently high intolerance level.

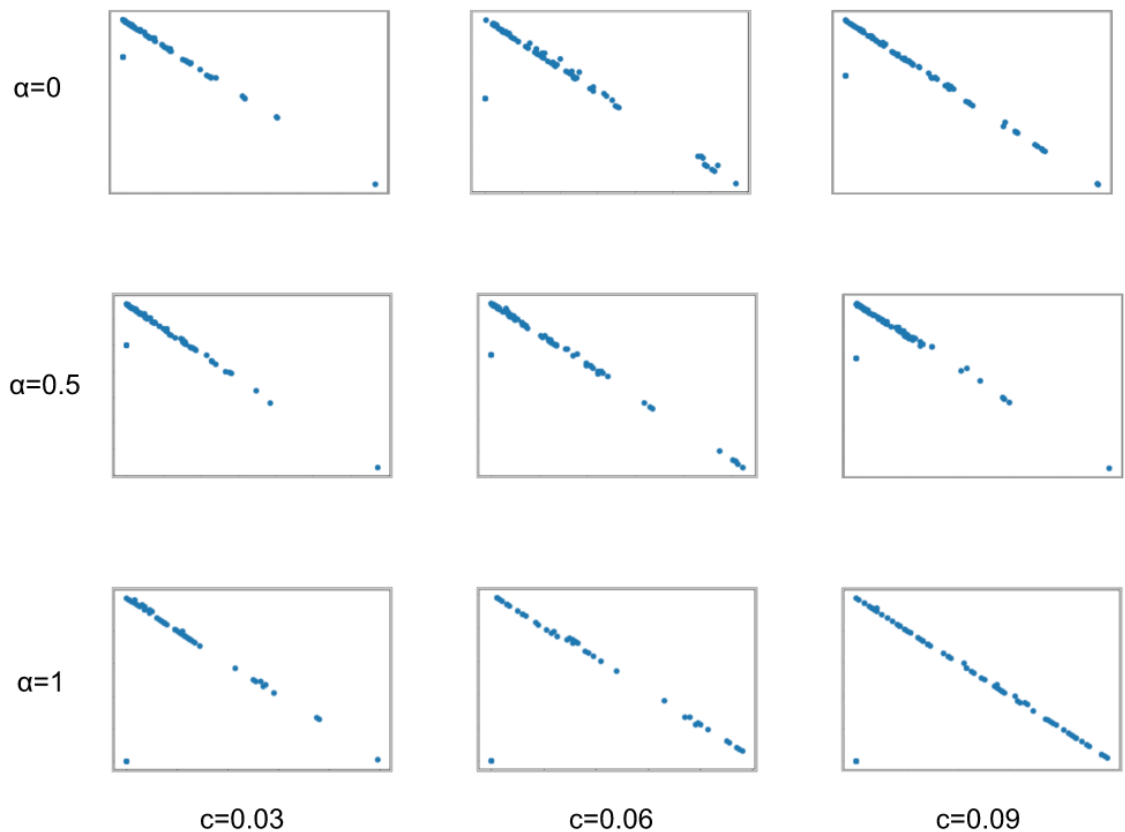

Figure 1: Scatter plots showing the overall distribution patterns in each scenario. The first row (α=0) represents high tolerance towards neighbors from different ethnic groups, the second row (α=0.5) is medium tolerance, and the third row (α=1) is high intolerance. The first column (c=0.3) represents low moving cost, the second column (c=0.6) is medium moving cost, and the third column (c=0.9) is high moving cost.

This finding shows that moving expenses can offset the effect of residential preferences on neighborhood segregation and sheds light on its role in social integration. Policymakers can benefit from using the same approach to investigate the impacts of other parameters in urban planning.

## 5. Conclusion

In this paper, RL is combined with ABM to simulate the residential outcome of a modified Schelling Segregation model. Segregation patterns under incentives such as neighbor preferences and moving expenses are presented. It is shown that high moving expenses promote social integration despite the level of intolerance towards neighbors from different ethnic backgrounds. The proposed simulation model can help policymakers study residential patterns and address challenges in areas such as urban planning. Further researchers can improve this model by exploring different RL algorithms or considering other realistic input parameters.